\documentclass[pre,floatfix,twocolumn,showpacs]{revtex4} 
\usepackage{graphicx} 
\usepackage{amssymb} 
\usepackage[mathscr]{eucal}

{\catcode`\|=\active \gdef|{\egroup\,\vrule\,\bgroup}} 
 
\newcommand{\Vec}[1]{\ensuremath \mbox{\boldmath $#1$}} 
 
\begin{document} 
\title{Optimal Path to Epigenetic Switching} 
 
\author{David Marin Roma} 
\author{Ruadhan A. O'Flanagan} 
\author{Andrei E. Ruckenstein} 
\author{Anirvan M. Sengupta} 
\affiliation{Dept.\  of Physics and Astronomy and  BioMaPS Institute, Rutgers University, Piscataway, NJ 08854} 
\author{Ranjan Mukhopadhyay} 
\affiliation{Dept.\  of Physics, Clark University, Worcester, MA 01610}

\begin{abstract} 
We use large deviation methods to calculate rates of  noise-induced transitions between states in  multi-stable genetic networks. We analyze 
a  synthetic biochemical circuit,  the toggle switch, 
and compare the results to those obtained from a numerical solution of the 
master equation.   
\end{abstract} 
 
\maketitle 
  
   
Fluctuations in bio-molecular networks have been the subject of much 
research activity recently \cite{rao}.  Studies on noise in gene 
expression \cite{elston, ozbudak, thattai, elowitz, swain, euknoise}, in signal 
transduction \cite{detwiler} and in biochemical oscillators \cite{barkai, 
vilar, gonze} demonstrated that having a small number of molecules affects, 
sometimes critically, the behavior of cellular circuits. Stochastic aspects of the 
choice between lytic and lysogenic developmental strategies of bacteriophage lambda virus 
infection in {\it E. coli} were studied in an influential  
paper by Arkin, Ross and McAdams \cite{arkin}. 

One of the interesting aspects developmental processes is that one could get multiple heritable cell fates without irreversible changes to the genetic information. Different cells with the same DNA sequence, showing different phenotypes that are stably maintained through cell divisions, namely epigenetic phenomena,
have been represented 
as  multiple stable attractors in deterministic descriptions of the 
biochemical dynamics. In this paper, we are concerned with 
the robustness of such attractors against spontaneous fluctuations which might induce transitions from one stable state to another.  
Previous work in this area has modeled the effects of fluctuations 
by adding Gaussian-distributed Langevin forces  to the deterministic 
equations \cite{bialek, sneppen, hasty}. Although this description is appropriate 
in describing typical fluctuations when the number of molecules is sufficiently large 
\cite{detwiler, elston, ozbudak, swain},  {\em rare events} involving occasional  
large departures from average behavior are typically outside the scope of the Langevin  
treatment (Gaussian approximation). The transition rate in a simplified model of the phage lambda switch has been studied \cite{sneppen, ao} in this approximation. We wish to compute the transition rate using a more appropiate large deviation theory with special focus on the attempt frequency. Of course, one could get the transition rate from direct computer simulations. However, direct simulations of rare events is, obviously, time-consuming. Recent research in the lambda switch  suggests that the simplified model lacks one very important physical interaction between distant regions of the lambda virus genome, changing dramatically the behaviour of the switch \cite{dodd}. Applying our tools to that question, among others, is the long time goal of our research. However, we wish to test our methods on a simpler system. We will consider the artificially constructed toggle switch \cite{toggle}. In this example, we find that the contributions to the transition rate coming from corrections to the Gaussian approximation can change the overall rate by several orders of magnitude and, therefore, are important for comparison with experimental results.

The theory of transition rates is a well developed subject (see \cite{hanggi} as well as references therein). For a bistable system like the genetic switch we are considering, the transition probability from one stable point to the other is estimated by computing the probability of reaching the saddle point between stable states, and, from there, to follow the deterministic trajectory to the other stable state, rather than  to fall back to the initial state. The transition rate is given by an expression of the form:
\begin{equation}\label{rate} 
\hbox{rate}=\frac{\lambda_+}{2\pi}\left[ \frac{detA_{fp}}{|detA_{sp}|}\right]^{1/2}*P(x_f,x_o), 
\end{equation} 
where $\lambda_+$ is the positive eigenvalue of the matrix describing the linearized equations of motion around the saddle point,  
$A_{fp}$ and $A_{sp}$ are the inverses of covariance matrices appearing in 
the quasi-stationary Gaussian approximation of the probability distribution in the starting 
stable point and in the saddle point respectively and $P(x_f,x_o)$ is the probability of finding the system at the saddle point state in a quasi-stationary distribution centered around the stable fixed point $x_o$. Note that $A_{sp}$ has one negative eigenvalue and the Gaussian distribution around the saddle point is only a formal solution. A more precise definition of $A$ appears later in the paper. For a derivations of a very similar formula see  \cite{langer} or section VII.D in the review \cite{hanggi}.

Much of the rest of the paper is devoted to the  computation of $P(x_f,x_o)$ by large deviation methods.  There are two related ways. In one approach,  one keeps track of the trajectories in the space of numbers of different molecules, distributed according to a state dependent Poisson process, and computes time dependent transition probabilities as sum the probabilities of all paths connecting the initial and final points, which leads naturally to a path integral formulation of the stochastic process. In this way, the transition probability is evaluated as the exponential of the ``action". This action can be computed in a perturbation expansion (using the volume of the system as a parameter), in which the leading order correction is the line integral along the path that minimizes the action (optimal path) of a Lagrangian function. This calculation naturally gives rise to a Hamiltonian that corresponds to the evolution operator in the master equation \cite{vk}, written in terms of numbers and raising and lowering operators expressed as exponentials of the phase variables conjugate to the numbers.

An alternative but exactly equivalent approach is to start directly from the master equation and solve it  in the Eikonal approximation \cite{dykman}. We will present our arguments in this article using this approach, which is easier to explain mathematically, and provides an easier way to compute the next order correction in the volume expansion, a term which has not been computed before in relation to these genetic switches. As we will see, in the case of the toggle switch,  the next order term in $\ln P(x_f,x_o)$ makes an important contribution to the overall rate of transition. 
   
The general ideas are developed in the context of the simple 
example of the toggle switch. This artificially realized switch 
consists of  two genes that repress each others' expression, placed in a high copy plasmid in {\it E. coli}. Once expressed, each protein can bind particular DNA sites upstream of the gene which codes for the other 
protein, thereby repressing its transcription.   
If we denote the $i$-th protein concentration by $x_i$, the deterministic system is described 
by the equations: 
\begin{eqnarray} 
\label{chem} 
\dot x_1 = \frac{a_1}{1+(x_2/K_2)^n}-\frac{x_1}{\tau}\\ \dot x_2 = \frac{a_2}{1+(x_1/K_1)^m}-\frac{x_2}{\tau} \end{eqnarray} 
the constants $a_1$ and $a_2$ incorporate all aspects of transcription 
and translation reactions. The Hill exponents, $m$ and $n$, represent the 
degree of cooperative binding of  proteins  to DNA, and $\tau^{-1}$ is the protein degradation/dilution 
rate (assumed equal for the two proteins). $K_1$  is the effective dissociation constant for binding of protein $1$ in the promoter of gene $2$.  $K_2$ is the corresponding parameter for  protein $2$.  
For some regions of parameter space, the system  has three stationary points: two stable ones and a saddle point \cite{toggle}. 
 
For the purposes of this discussion, we model the stochastic evolution of the protein concentrations in the system by a birth-death process in which protein $i$ is made in short-lived bursts of 
size $b_i$ and proteins  are diluted or degraded at a rate 
$\tau^{-1}$. A more detailed description involving proteins and RNA 
will be published elsewhere. It is worth noting that, while both the burst size $b_i$  
and the RNA production rate show up as parameters in the stochastic modeling, only their product,  
$a_i$,  shows up in the effective deterministic equations (\ref{chem}) for the protein levels.  
 
To compute the rate of transition from one fixed point to the other, 
we must solve the master equation \cite{vk}, which describes the time evolution 
of the probability distribution of protein concentrations. The qualitative behavior of the 
stationary solution for the bistable system can be described in simple intuitive terms: 
the solution displays two peaks 
centered around the stable points. If we start with probability one 
around one of the stable points, rare transitions lead to a long tail which leaks into the domain of attraction of the other stable point, in very much the same way in which the probability amplitude extends beyond the classically allowed region in quantum mechanical tunneling through a barrier. This analogy motivates the Eikonal approximation to the solution of the master equation 
\cite{dykman}. The master equation is given by, 
\begin{equation} 
\frac{\partial P}{\partial t} =\Omega\sum_e \lbrack  W_{\hat e}(\Vec x - \hat e/\Omega)P(\Vec x -\hat e/\Omega,t) - W_{\hat e}(\Vec x )P(\Vec x ,t) \rbrack 
\label{master} 
\end{equation}
where $\Omega$ is the volume of the system,  
$\hat e/\Omega=\Delta \vec x$  
is the 
concentration change associated with individual reaction events, the rate of which is given by $\Omega W_{\hat e}(\Vec x )$. 
Assuming that the distribution is quasi-stationary in 
the region of interest, we consider solutions of the WKB form: 
\begin{equation} 
\label{wkb} 
P(\Vec x,t) = C\exp\lbrack-\Omega S(\Vec x),\rbrack  \hspace{0.5cm}	S(\Vec x_{o})=0. 
\label{prob} 
\end{equation} 
$x_{o}$ being the initial stable point. 
In the same way the wave function in quantum mechanics is computed using an expansion in powers of $\hbar$, it customary to find the probability $P(\Vec x,t)$ by expanding $S(\Vec x)$ in powers of inverse volume, which plays the same role as $\hbar$ in quantum mechanics, since the bigger the volume, the less likely are fluctuations to happen. Then, to first order in $\Omega^{-1}$, we write: \[S(\Vec x)=S_0(\Vec x) +\Omega^{-1} S_1 (\Vec x) +O(\Omega^{-2}).\]
Assuming that the scaled transition rates $W_{\hat e}(\Vec x )$ are smooth functions of 
$\Vec x$, and expanding $S$ to first order,  $S(\Vec x - \hat e/\Omega) 
= S(\Vec x) -  \frac{\hat e_i}{\Omega} . \frac{\partial}{\partial x_i} 
S(\Vec x)$, collecting the terms which do not contain powers of $\Omega$ we have: 
\begin{eqnarray} 
\frac{\partial P(\Vec x,t)}{\partial t} = H P(\Vec x,t) \\ 
H(\Vec x,\Vec p) =\displaystyle  \sum_{\hat e} \lbrack  W_{\hat e}(\Vec x)(e^{(\hat e  .  \Vec  p ) } - 1) \rbrack
\label{hdef} 
\end{eqnarray} 
where  $H$ is the Hamiltonian describing the time evolution of the 
probability distribution, and we define the momentum $p_i$ as:
\begin{equation}
\label{momentum}
 p_i = \frac{\partial}{\partial x_i} S_0(\Vec x)
\end{equation}

 If we expand the Hamiltonian \ref{hdef} in $\Vec p$ and keep terms up to second order in $\Vec p$ we recover the Gaussian 
approach used in \cite{bialek, sneppen}. Since we are considering a situation where the transitions are so rare that the probability does not change much in time, the Hamiltonian will be very small. 

The main contribution to the transition probability 
is obtained by evaluating $P$ along a particular trajectory \cite{dykman}. This trajectory, called the optimal path, is the solution to Hamilton's equations derived from  Eq.\ref{hdef}:

\begin{small}  
\begin{eqnarray} 
\label{optimalpath}
\dot x_i = \frac{\partial H(\Vec x,\Vec p)}{\partial p_i}=\sum_{\hat e} \lbrack \hat e _i W_{\hat e}(\Vec x)e^{(\hat e_a  p_a )}  \rbrack \\  
\dot p_i =-\frac{\partial H(\Vec x,\Vec p)}{\partial x_i}= 
-\sum_{\hat e} \lbrack \frac{\partial W_{\hat e}(\Vec x)}{\partial x_i}(e^{(\hat e_a p_a)}  - 1) \rbrack 
\label{peqn}
\end{eqnarray} 
\end{small}

For the toggle switch example we have four  $\hat e_i$-s describing jumps to the right, left, up or down, given by $b_1\hat x_1, -\hat x_1, b_2\hat x_2, \textrm{ and } -\hat x_2$, respectively. 
The relevant Hamiltonian defined on times long compared to the inverse binding/unbinding rates of proteins at the two promoters is given by:
\begin{widetext}
\begin{equation} 
\label{Hamiltonian}
H=  \frac{ a_1/b_1}{(1+(x_2/K_1)^n)}\left( e^{b_1p_1}-1\right)  + \frac{ x_1}{\tau}\left( e^{-p_1}-1\right) +  \frac{
a_2/b_2}{(1+(x_1/K_2)^m)}\left( e^{b_2 p_2}-1\right)  +  \frac{x_2}{\tau}\left( e^{-p_2}-1\right) .  \end{equation}
\end{widetext}
As already mentioned above, $K_{1,2}$ are the effective dissociation constants for binding of proteins $1,2$ at the promoter of gene $2,1$, respectively, $b_i$ is the burst size of protein $i$ and the ratio $a_i/b_i$ is a measure of the RNA
production rate associated with the transcription of the gene $i$.

To extract the values of the burst size parameters, the  spontaneous transition rate has to measured experimentally for more than one conditions. Since this has not yet been done,  we will compare the results of the Eikonal approximation to the solution obtained by direct diagonalization of the Hamiltonian (\ref{Hamiltonian}). For simplicity we will set the parameters $K_i=1$,$b_i=1$.
 
\begin{figure}[!h] 
{\centering 
\includegraphics[height = 9cm,angle=-90]{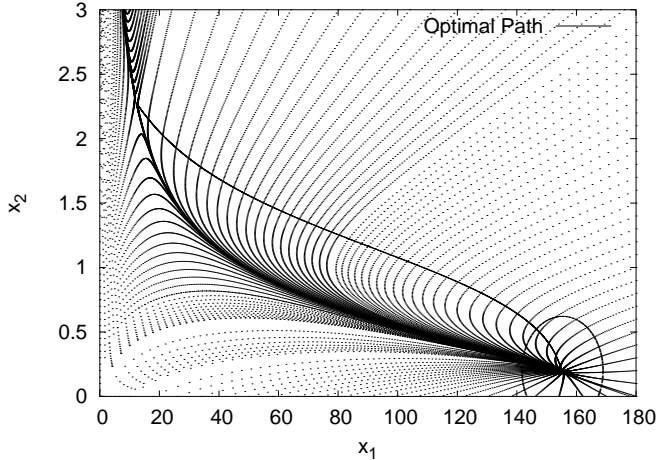} 
\caption{Optimal path for the  parameters, $a_1=156$, $a_2=30$,$n=3$,$m=1$, $K_1=K_2=1$, $b_1=b_2=1$ and $\tau=1$. 
$xi$ are dimensionless. The ellipsoid  indicates the orientation
of the Gaussian spread  around the stable point.  The size of the spread  scales like $\Omega^{-\frac{1}{2}}$.
} 
\label{fig:path} 
 \par} 
 \end{figure}

The optimal path for the transition from one stable point to the other 
starts near one stable point, proceeds to the saddle point and from there it follows 
the deterministic trajectory to the other stable point.  
Thus we must first find solutions of Eqs. \ref{optimalpath} and \ref{peqn} which start at (near) the initial stable point and end at the saddle point. At the end points we have $p_1=p_2=0$,  and $H=0$.  
This also implies that if the system is at the stable point  it  will remain there.
So, the optimal path must instead start at a point very 
close to but not exactly at the fixed point. In this case, the Hamiltonian will be a very small number (and constant). In what follows, we will make the approximation $H=0$. The initial conditions for the momentum equations can be 
obtained by approximating the probability around the stable point by 
a Gaussian distribution $P=e^{-{\Omega S_g}}$ with  
$S_g=\frac{1}{2}A_{ij}\delta x_i\delta x_j$ (note that we use summation convention, i.e., repeated indices are summed over). Then   
$p_i=\frac{\partial S_o}{\partial x_i}=A_{ij}\delta x_j$,  
and we expand the equation $H=0$ around the stable point to find $A_{ij}$. Then we have a two point boundary value problem which can be solved by various methods \cite{numrec}.   
The solution of the equations of motion \ref{optimalpath} and \ref{peqn} for a set of parameters, projected to concentration space, is shown in Figure \ref{fig:path}. We integrate equations \ref{momentum} along the optimal path C  to obtain $S_0=\int_C p_idx_i $ .

The $S_1$ factor can be viewed as a correction due to fluctuations around the optimal 
path and could be calculated following references \cite{onedim, arizona}. Collecting coefficients of powers of $\Omega$ in the $\Omega^{-1}$ expansion  we  derive an equation for $S_1$:
  
\begin{small} 
\begin{equation}  
\label{s10} 
\sum_{\hat e}\lbrack W_{\hat e} \hat e_i\frac{\partial 
S_1}{\partial x_i} -  \frac{W_{\hat e}}{2}\hat e_i \hat e_j\partial_ip_j - 
\hat e_i\partial_iW_{\hat e}\rbrack e^{(\hat e_a  p_a )}=0 \end{equation} 
\end{small}  
In turn, after using the equations of motion to rewrite the first term as derivative along the optimal path $x_{op}(t')$, Eq. (\ref{s10}) can be transformed into: 
 \begin{small} 
\begin{equation}\label{s1} 
\frac{d}{dt'}S_1 = \sum_{\hat e}  \frac{1}{2}W_{\hat e} \left(\Vec 
x\right)\hat e_i \hat e_j \frac{\partial 
p_j}{\partial x_i} e^{\hat e_i p_i}
+\sum_{\hat e}\hat e_i\frac{\partial 
W_{\hat e}\left({\Vec x}\right)}{\partial x_i}e^{\hat 
e_i p_i} \end{equation} 
\end{small} 
 
To proceed we need $\frac{\partial p_j}{\partial x_i}$ along the path. From 
Hamilton's equations (\ref{optimalpath}) it follows that $\delta 
p(t)_a = M(t)_{ab}\delta x(t)_b$, and thus we can use the components of the matrix {\it M}  
in place of the derivative $\frac{\partial p_j}{\partial x_i}$ in (\ref{s10}). Moreover, (\ref{optimalpath}) also  
implies that:  
\begin{eqnarray} \delta \dot { x^a} =  \frac{\partial^2 H}{\partial p_a \partial x^i} \delta 
x^i  +	\frac{\partial^2 H}{\partial p_a\partial p_i} \delta p_i \\ \delta \dot { p_a} = 
- \frac{\partial^2 H}{\partial  x^a \partial x^i} \delta x^i  -  \frac{\partial^2 H}{\partial  x^a \partial 
p_i} \delta p_i \end {eqnarray}
Combining this together with the time derivative of $\delta p(t)$, 
\begin{equation} \delta \dot p = \dot M \delta \Vec x + M \delta \dot {\Vec x}  
\end{equation} 
leads to the following set of coupled differential equations for $M$: 
\begin{eqnarray} 
\label{M-eq} 
\dot M_{ab}&+& M_{ac}\frac{\partial^2H}{\partial x^b \partial p_c}+M_{ac}\frac{\partial^2H}{\partial p_c 
\partial p_d}M_{db} \nonumber\\
&+&\frac{\partial^2H}{\partial x^a \partial p_c}M_{cb}+\frac{\partial^2H}{\partial x^a \partial x^b}=0\\
& &\nonumber 
\end{eqnarray} 
with initial conditions: $M_{ij}(t=0)=A_{ij}$ (defined below equation \ref{Hamiltonian}).   Finally, solving these equations together with 
equations \ref{optimalpath} and \ref{peqn} we integrate equation (\ref{s1}) to obtain $S_1$. 
Given the above values of $S_0$ and $S_1$ we compute the transition probability, $P(x_f,x_o)$,
from the starting stable point, $x_o$,  to the saddle point, $x_f$. Using equation \ref{rate} we can, therefore, find the transition rate for any large value of $\Omega$. We now compare this calculation to the direct estimation of transition rates as described below.

From the  master equation (\ref{master}), it follows that the eigenvalues of $H$ measure the decay rates  of  non-stationary states corresponding to eigenvectors of $H$ with  nonzero eigenvalues. The equilibrium state is represented by the ``zero mode", i.e., the eigenvector of $H$ with zero eigenvalue, the existence of which is guaranteed by the transition matrix character of the Hamiltonian and conservation of probability. 
To compute the eigenvalues of the Hamiltonian, we write the master
equation in discrete form, replacing the continuous concentration variables $(x_1,x_2)$ with
a lattice with lattice parameter $1/\Omega$. Although the system displays infinitely many
states, typically, the gap between the real parts of the eigenvalues for first and second excited states is much larger than the absolute value of the real part of the first eigenvalue. This is because the gap between the first excited state and the second or the third excited states are governed by local relaxation rate around the two fixed points, but, the gap between the ground state and the first excited state is governed by  the transition rate between the two stable fixed points. The local relaxation rates are order one in $\Omega$, whereas, the transition rate  is exponentially small for large $\Omega$ (in practice, we find the ratios of the real parts to be about $10^3$).
Thus an arbitrary probability distribution  rapidly
decays into a linear combination of the stationary state and the first
excited state. Equivalently, the state could be described as a linear
combination of two states, each representing a quasi-stationary
distribution around a stable fixed point. From then on,
we can project the evolution to this  two state system. If we start with
probability $p_o$ of being in the state $(1,0)^T$, 
then the Master equation gives:

\[ \frac{d}{dt}\left( \begin{array} {c} p_o\\ p_f \end{array}\right)= \left( \begin{array} {cc} -r_{12}& r_{21}\\ r_{12}&
-r_{21}\end{array}\right) \left( \begin{array} {c} p_o\\ p_f \end{array} \right)
 \]

The two-by-two effective transition matrix has columns which sum to zero
ensuring probability conservation. Also, the trace $ 0 +
\epsilon_1 = r_{12} + r_{21}$, where $\epsilon_1$ is the eigenvalue
of the first excited state.  Therefore the first excited eigenvalue
will be the sum of the forward and backward rates. In the case of the
asymmetric systems, one rate is usually far greater than the other. Consequently
the larger rate among $r_{12}$ and $r_{21}$ will be
approximately given by $\epsilon_1$, which we computed numerically
using the Matlab routine  ``eigs" for sparse matrices as well as by Lanczos algorithm \cite{lanczos}. For a symmetric choice of parameters for the
two proteins, each rate is just $\epsilon_1/2$.

\begin{figure}[!tbp] 
\includegraphics[height = 9cm,angle = -90]{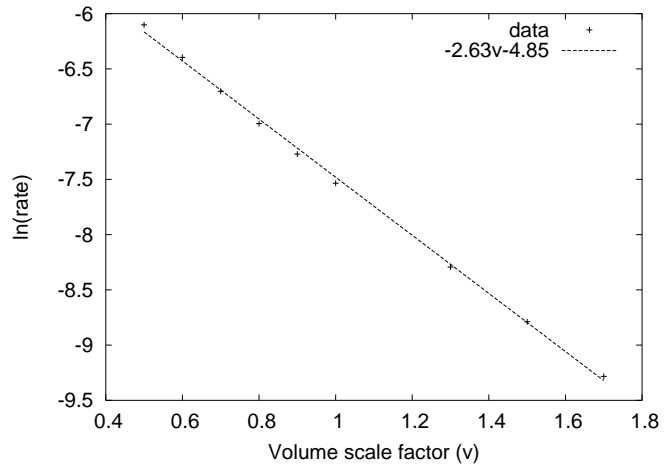} 
 \caption{Scaling with volume:
  estimates from direct computation of eigenvalues are
  $S_0 = 2.63$, $S_1+\ln(\text{pref})=4.85$ whereas optimal path calculation  gives $S_0 = 2.47$,
	$S_1= 3.5$, $\ln(\text{pref}) = 1.5$. In this example the backward rate is 1000 times smaller than the forward rate, so the lowest nonzero eigenvalue is very close to the rate of switching.}
   \label{volume}
  \end{figure}

To explicitly extract the $S_0$ and $S_1$ contributions to the rate from the Lanczos results, we re-scale the volume of the system $\Omega \to \nu \Omega$ which, in turn, leads to a re-scaling of rates of individual reaction events as $f(x)\rightarrow \nu f(x)$.  
As a function of volume scale factor, $\nu$, the logarithm of the rate has the form:
$ln(r)= S_0\nu + b$, where $b$ includes both  $S_1$ and the logarithm of the pre-factor of $P(x_f,x_o)$ in Eq.\ref{rate}. The results and comparison with the Eikonal approximation are shown in Fig.\ref{volume}. The dotted line is a fit to the data points obtained from calculation of the eigenvalues, and we see that the slope and intercept computed from equations \ref{rate},\ref{prob} are in good agreement with these values. Note that, in this example,  $S_1$ and the pre-factor are significant contributions to the transition rate. 

When we perform these calculations for the ``standard" model of  the lambda switch \cite{sheaackers, sneppen}, we find a rate three orders of
magnitude higher than the observed rate of $10^{-7}$ per generation \cite{little}. In retrospect, it is clear that accounting for the stability of the lysogenic state requires a more complex model which should include the effect of DNA looping \cite{dodd}.  Whether the stability is due to suppression of fluctuation or due to disappearance of the lytic ``fixed" point \cite{santillan} remains an open question.

Optimal path methods are routinely used for studying rare events
related to failure of communication networks modeled as birth and death
processes \cite{communications}. Such large deviation
methods are likely to be important in the context of robustness and adaptability of
biological networks. This paper illustrates the power
of an approach to fluctuations based on the Eikonal approximation to solutions of the master equation. The scheme  incorporates large deviations in a natural way and provides a quantitative method scalable to large networks. 
We also hope that beyond being an efficient computational tool, 
this method will provide further insight into to the
stability of epigenetic states of complex genetic networks.
			 
Acknowledgment:  AMS thanks John Little and John Reinitz for useful discussions. AER and AMS acknowledge the hospitality of the  Kavli Institute of Theoretical Physics and of Aspen Center for Physics while this work was being done. This work was supported in part by an NIH P20 Grant GM64375.
 
 \end{document}